\documentclass[a4paper,fleqn,usenatbib,useAMS]{mnras}

\usepackage{graphicx}   
\usepackage{amsmath}    
\usepackage{amssymb}    
\usepackage{multicol}        
\usepackage{bm}         
\usepackage{pdflscape}  

\usepackage[T1]{fontenc}
\usepackage{ae,aecompl}

\usepackage{newtxtext,newtxmath}

\title[Flow Properties of XTE~J1752-223]{Inference on Accretion Flow Properties of XTE~J1752-223 During Its 2009-10 Outburst}
\author[Chatterjee et al.]{Kaushik Chatterjee$^1$, Dipak Debnath$^1\thanks{E-mail: dipakcsp@gmail.com}$, Debjit Chatterjee$^1$, Arghajit Jana$^1$, 
\newauthor{Sandip K. Chakrabarti$^{1}$}\\
$^1$ Indian Centre For Space Physics, 43 Chalantika, Garia Station Road, Kolkata, 700084, India}

\date{Accepted 2020 February 6. Received 2020 January 25; in original form 2019 May 6}

\pubyear{2020}

\begin{document}
\maketitle

\begin{abstract}

We carry out a detailed study of the spectral and the timing properties of the stellar-mass black hole candidate XTE J1752-223 
during its 2009-10 outburst using RXTE PCA data in the $2.5-25$~keV energy range. Low frequency quasi-periodic oscillations 
(LFQPOs) are seen in the power density spectrum (PDS). The spectral analysis is done using two types of models: one is the 
combined disk black body plus power-law model and the other is Transonic flow solution based Two Component Advective Flow (TCAF) 
model. RXTE PCA was non-operational during 2009 Nov. 16 to 2010 Jan. 18 and thus we study light curve profiles and evolution 
of hardness ratios using MAXI GSC and Swift BAT data. Based on the evolution of the temporal and the spectral properties of the 
source during its 2009-10 outburst, we find that the object evolved through the following spectral states: hard, hard-intermediate 
and soft-intermediate/soft. From the TCAF model fitted spectral analysis, we also estimate  the probable mass of the 
black hole to be in the range of $8.1-11.9$~$M_\odot$, and more precisely, the mass appears to be is $10\pm{1.9}~M_\odot$. 

\end{abstract}


\section{Introduction}

Compact objects, such as, neutron stars and black holes (BHs) are the end products of massive stars. The gravitational pull of BHs 
is so high that no particle or radiation can escape from them. They both can be detected only by the electromagnetic radiation emitted 
by the accreted matter falling on them. Most of the black hole candidates (BHCs) in our Galaxy are in close binaries with companion 
stars which act as donors. Wind material from the companion star or matter accreting via Roche lobe overflow falls towards the black 
hole due to its intense gravitational pull and starts swirling around it to form accretion disk due to the presence of angular momentum. 
The observed electromagnetic radiation from a black hole binary is from the accretion flow itself. Some of the black hole candidates in 
low mass X-ray binaries are transients in nature, which occasionally undergo outbursts. The increased X-ray flux in these transient X-ray 
binaries shows variability in the temporal and spectral states. Several works have already been done on this and many papers are available 
in the literature (see, e.g., Tomsick et al. 2000; McClintock \& Remillard 2006; Debnath et al. 2008, 2013; Nandi et al. 2012; Rao 2013) 
to explain the variation of spectral and temporal properties of these objects during their active X-ray outburst phases. It is also reported 
by some authors that these objects transit through several spectral states (e.g., hard state (HS), hard-intermediate state (HIMS), 
soft-intermediate state (SIMS), soft state (SS)) during their active outburst phases (see Debnath et al. 2013 and references therein). 
Depending upon the observed spectral states, Debnath et al. (2017) classified these transient BH binaries into two types: classical or 
type-I (all four states are observable) and harder or type-II (softer states are missing). Low and high frequency quasi periodic oscillations 
(QPOs) are generally observable in the power density spectra (PDS) of these sources (see, Remillard \& McClintock 2006 for a review). Low 
frequency QPOs are commonly observed in hard and intermediate spectral states. Generally, it has been observed that during the rising HS and 
HIMS, frequency of these QPOs monotonically increases with time (day) and during the declining phase, the opposite nature is observed. In 
SIMS, these QPOs are observed sporadically. 

It is well-known that the emitted spectrum, emerging from black hole candidates (BHCs), primarily contains a combination of a multi-color 
thermal blackbody component and a non-thermal power-law component. The thermal component is the radiation coming from the standard Keplerian 
disk (Shakura \& Sunyaev 1973) and the non-thermal component originates from the hot `$Compton$' cloud (Sunyaev \& Titarchuk 1980, 1985). 
The standard disk contains soft X-ray photons. The `$Compton$' cloud is the repository for hot electrons. The energy of these hot electrons 
is transferred to those soft X-ray photons in the process of repeated inverse Compton scattering to produce high-energy X-ray photons and 
this gives the hard power-law tail in the observed spectra of an accreting BH. In TCAF solution (Chakrabarti 1995, 1997; Chakrabarti \& 
Titarchuk 1995, hereafter CT95), the `$hot~corona$', which is known as the `$Compton~cloud$', is the CENtrifugal pressure supported 
BOundary Layer (CENBOL) that forms behind the centrifugal barrier due to the piling up of the low viscous and low angular momentum optically 
thin matter, known as the sub-Keplerian (halo) component. Another component in TCAF solution is the optically thick and geometrically thin 
Keplerian (disk) component which is submerged inside the sub-Keplerian component. In 2014, this TCAF solution has been implemented into 
HEASARC's spectral analysis software package XSPEC as an additive table model after generation of the model {\it fits} file to fit 
energy spectra of black holes (Debnath et al. 2014, 2015a). Now, one can extract information about the physical flow parameters directly 
from the spectral fits (Debnath et al. 2014, 2015a,b, 2017; Mondal et al. 2014, 2016; Jana et al. 2016, 2019; Chatterjee et al. 2016, 2019). 
Estimation of the intrinsic source parameters such as the mass of the black hole (Molla et al. 2016, 2017; Debnath et al. 2017 and references 
therein; Nandi \& Chakrabarti 2019) and X-ray flux contributions from jets/outflows are also possible (Jana et al. 2017, 2019; Chatterjee et al. 
2019) from spectral analysis with the TCAF model.

The stellar-mass black hole (SBH) candidate XTE J1752-223 was monitored and discovered by Rossi X-ray Timing Explorer (RXTE) on  October 21, 
2009 (Shaposhnikov et al. 2010) at the position R.A.=$268.05\pm0.08$, Dec=$-22.31\pm0.02$ (J2000 coordinate system). Markwardt et al. (2009a) 
proposed the source as a BHC based on their analysis. Ratti et al. (2012) suggested the location of the source to be in the Galactic bulge. 
During the outburst, the source was active for almost 8 months. RXTE monitored the source roughly on a daily basis except from Nov. 20, 2009 
to Jan. 19, 2010 as during this phase RXTE pointed observations were not possible due to the Sun constraint. After the discovery the source 
was also monitored by MAXI, Swift and other satellites. The X-ray behavior of the source during the outburst was matched with typical 
phenomenological picture of the black hole transients, confirming XTE~J1752-223 as a strong accreting black hole candidate (Nakahira et al. 
2010; Shaposhnikov et al. 2010; Curran et al. 2011). Strong, relativistic iron emission lines were detected by $Suzaku$ and $XMM-Newton$ 
(Reis et al. 2011). Reis et al. (2011) estimated spin of the source as $a=0.52\pm0.11$. Miller-Jones et al. (2011) estimated the 
disk inclination angle of the source to the line of sight as $i < 49^\circ$. Torres et al. (2009a,b) reported a bright optical and a 
near-infrared counterpart just after the X-ray discovery. Brocksopp et al. (2009) reported a radio counterpart with a flux density of 2 mJy 
at both 5.5 and 9 GHz using Australia Telescope Compact Array (ATCA) with a spectrum consistent with that of a compact jet in hard state. 
Brocksopp et al. (2010b,c) showed the presence of resolved jet components by additional radio observations using the European VLBI Network 
(EVN). In the intermediate state the radio emission still produced compact jets and the radio flux became more variable (Brocksopp et al. 2013).
Russell et al. (2012) discussed a late jet re-brightening in the decaying hard state from multi-wavelength observations. Shaposhnikov et al. 
(2010) found that the source exhibited all the canonical spectral states of a classical black hole candidate. They also 
reported a mass estimate for the black hole $M_{BH}=9.8\pm0.9~M_\odot$ and  a distance of $d \sim 3.5\pm0.4$ kpc. According to Brocksopp
 et al. (2013), this object was jet active during the entire duration of the outburst. According to their report, the jet was compact till 2010 
Jan. 21 (MJD 55217.87) on which the 1st peak in the radio light curve (both 5.5 and 9 GHz) occurred. Again, the jet became compact after 2010 March 
24 (MJD 55279.63) where the last radio peak occurred. In the intermediate region, optically thin radio flares were present which gradually decreased 
in flux density. Shaposhnikov et al. (2010) reported the existence of high soft state during this outburst from MJD $\sim 55220.68$ to MJD 
$\sim 55279.63$ when high radio flares were active (Brocksopp et al. 2013). It is reported in the literature that compact jets 
are generally seen in hard states, while discrete or blobby jets are observed in intermediate (HIMS and SIMS) states (see, Jana et al. 
2017 and references therein). In soft states, jet bases are cooled down and no jet is observed. In the present case, the source has a 
short orbital-period ($P_{orb} \leq6.8$ hr) with an ${\textit{M}}$ type mass donor as the companion star (Ratti et al. 2012). 
Recent study of this types of objects (MAXI~J1659-152, MAXI~J1836-194, Swift~J1753.5-0127, XTE~J1118+480) by our group suggests that 
these type-II or harder type of outbursting sources do not show soft spectral state (Debnath et al. 2015b, 2017; Jana et al. 2016; 
Chatterjee et al. 2019). In this paper, we intend to study the accretion flow properties during this outburst through careful analysis 
of the evolution of the spectral and the temporal properties with two types of models: phenomenological disk blackbody plus power-law (DBB+PL) 
models and physical TCAF model. Through these analysis we wish to verify if this short period source indeed has a soft state during this outburst 
as reported by previous authors. If the answer is `yes' then our goal would be to find physical mechanisms for the production of jets in the soft state.

This paper is organized in the following way. In \S 2, we briefly discuss about the data analysis technique using HEASARC's HEASoft 
package. In \S 3, we present spectral and temporal analysis results of the source. Finally, in \S 4, a brief discussion on the results of 
the accretion flow dynamics of this BHC and concluding remarks are presented.

\section{Observation and Data analysis}

RXTE PCA monitored XTE~J1752-223 roughly on a daily basis (except operational shutdown during 2009 Nov. 16 to 2010 Jan. 18) till 2010 Aug 
6 after its discovery on October 2009. Here we analyzed $32$ observations of RXTE PCA from October 26, 2009 (MJD 55130.96) to June 21, 2010 
(MJD 55368.94) to study the spectral and timing properties of this source during this outburst. We have used HEASARC's software package 
HEASoft version HEADAS 6.19 and XSPEC version 12.9 for the analysis of the archival data of RXTE PCA instrument. We extract data 
from the best calibrated proportional counter unit 2 (PCU2). We generally follow the methods mentioned in Debnath et al. (2013, 2015a) for 
data reduction and analysis.    

For timing analysis, we also create the power density spectra (PDS) which are generated using XRONOS task `powspec' on  $0.01$ sec time binned
RXTE PCA light curves in the energy range $2-15$~keV to find any signature of low frequency quasi periodic oscillation (QPOs). The normalization
factor in `powspec' is taken to be `-2' to have expected `white' noise subtracted rms fractional variability. The centroid frequency of the 
QPOs is found by fitting the PDS with Lorentzian profiles. The coherence parameter $Q~(=\nu/\bigtriangledown\nu)$ and amplitudes (= $\%$ rms)
make the selection of QPOs in PDS where, $\nu$, $\bigtriangleup\nu$ are the centroid QPO frequency and the full-width at half-maximum 
respectively as discussed in Debnath et al. (2008). We also used RXTE ASM light curve data to study the evolution of dynamic photon index ($\Theta$) 
and variation of intensity (1.5-12 keV) with it to study the temporal properties. The dynamic photon index ($\Theta$) is the slope of the
hard region of a spectrum (Ghosh \& Chakrabarti (2019). We have discussed about $\Theta$ in more detail in the result section (3.1.2).
We also studied MAXI GSC (in 2-10 keV) and Swift BAT (in 15-50 keV) daily average light curves and their count ratios for our temporal study.  

For spectral analysis, the standard data reduction procedure for extracting the RXTE PCA (PCU2) spectral data are used. The background 
subtracted spectra are first fitted with the phenomenological disk black body ($DBB$) + power-law ($PL$) model. We have fitted $32$ PCA 
observations with the $DBB+PL$ model in the energy range of $2.5-25$~keV. A fixed value of 1$\%$ systematic error and hydrogen column 
density ($N_H$) of $0.46 \times 10^{22}~atoms~cm^{-2}$ (Shaposhnikov et al. 2010) for photoelectric absorption model $phabs$ are used 
to fit the spectra. Additional Gaussian iron line of peak around $6.4$~keV for $Fe$~emission line is used to achieve the best fit. The 
fluxes of different model components of the fitted spectra are calculated using the $cflux$ calculation method. 

We refit all the spectra with the current version (v0.3.1) of TCAF model $fits$ file as an additive table model into HEASARC's 
spectral analysis software package XSPEC. Here a fixed value of 1$\%$ systematic error also has been used. For fitting the spectra of a 
black hole (BH) with TCAF model {\it fits} file, one requires to supply six parameters including the mass of the BH and normalization. The 
normalization is expected to remain constant or vary in a very narrow range throughout the entire outburst unless there is an occurrence of 
phenomenon like jets (for more details see, Molla et al. 2016; Jana et al. 2017). The model input parameters are: $i)$ Keplerian disk rate 
({$\dot{m}_d$ in Eddington rate ${\dot{M}}_{Edd}$), $ii)$ sub-Keplerian halo rate ({$\dot{m}_h$ in ${\dot{M}}_{Edd}$), $iii)$ location of 
shock ($X_s$ in Schwarzschild radius $r_s=2GM_{BH}/c^2$), $iv)$ compression ratio ($R=~{\rho}_+/{\rho}_-$, where ${\rho}_+$ and ${\rho}_-$ 
represent the densities in the post-shock and pre-shock flows), $v)$ mass of the BH ($M_{BH}$) in solar mass ($M_{\odot}$) unit. The normalization 
is a function of the distance and inclination angle. Unless there is any separate dominating physical processes whose effects are not considered 
in the present version of the TCAF model {\it fits} file, or disk is not precessing, the normalization should not vary from observations to 
observations. Our procedure is to keep it as a free parameter while fitting the spectra, and it generally comes into a narrow range throughout 
the outburst(s) of a particular BHC observed with same satellite instrument (see, Molla et al. 2016, 2017). This would determine the normalization 
once and for all even without knowing the distance and inclination angle. In case the mass of the source is also not known, one may also keep 
it as a free parameter to find probable value of it (see, Molla et al. 2016; Debnath et al. 2017 and references therein) from each observation 
using this normalization constant. To obtain the best model fit (${{\chi}^2}_{red} \simeq 1$), a Gaussian iron line of peak around $6.4$~keV is 
used. In some of the observations another additional Gaussian line of peak around $4.2$~keV is used, which may be due to red-shifted $Fe K_\alpha$ 
line. 

\section{Results}

Studying X-ray properties in both temporal and spectral domains of transient black hole candidates helps understanding the accretion flow properties 
of these sources during various outbursting phases. Here, we report both these properties of the Galactic BHC $XTE~J1752-223$ during its 2009-10 
outburst. 

The detailed analysis results are given in Table 1, 2. In Table 1, primary (dominating) observed QPO parameters obtained from Lorentzian model 
fitted are mentioned. In Table 2, we mentioned spectral parameters obtained from fits using two types of models: i) combined DBB and PL models, 
ii) TCAF model based {\it fits} file in XSPEC.

In Fig. 1 (a-b), variation of MAXI GSC (in 2-10 keV; online red), Swift BAT (in 15-50 keV; online blue) light curves and hardness
ratios (BAT/GSC counts; online green) are shown. In Fig. 1 (c) the variation of radio light curves at both 5.5 and 9 GHz frequencies are shown. 
These radio light curve data are collected from Brocksopp et al. (2013). In Fig. 2, we show the hardness intensity diagram (HID). In Fig. 3 (a-b), 
we show the variation of dynamic photon index ($\Theta$) and variation of intensity with $\Theta$. In Fig. 4 (a-c), we show the variations of PCA 
count rate, total flux (DBB+PL) and total rate (${\dot m}_d + {\dot m}_h$) with day in MJD. In Fig. 5 (a-d), the variations of $DBB+PL$ fitted 
power-law flux, disk black body flux, power-law photon index ($\Gamma$) and disk temperature ($T_{in}$) with MJD are shown, which are obtained 
by fitting PCA spectra in $2.5-25$~keV energy range. Similarly in Fig. 6, we show the variations of TCAF model fitted parameters: disk rate 
(${\dot m}_d$), halo rate (${\dot m}_h$), shock location ($X_s$), compression ratio ($R$) and mass of the source ($M_{BH}$). In Fig. 7, accretion 
rate ratio intensity diagram (ARRID) is shown.

\subsection{Temporal properties} 

To study temporal behaviour of the source during its 2009-10 outburst, we studied evolution of dynamic photon index ($\Theta$) during the entire 
course of the outburst and also studied the variation of intensity with $\Theta$ using the RXTE ASM data. We also studied light curves using 
combined MAXI GSC and Swift BAT as well as RXTE PCA. Evolution of $\Theta$ and hardness ratio is studied to understand rough nature of the spectral 
states during different phases of the outburst. We also studied the fast-Fourier transformed PDS (done using 0.01 sec time binned PCU2 light curves) 
to search for low frequency QPOs and understand their properties.

\subsubsection{Light Curve Evolution}

During the rising phase of the outburst, due to scheduled maintenance and sun constraint RXTE PCA was non-operational during 2009 Nov. 16 (MJD=55151) 
to 2010 Jan. 18 (MJD=55214). So, to study outburst profile and evolution of hardness ratio, we used daily average light curves of MAXI GSC in $2-10$
~keV band and Swift BAT in $15-50$~keV band (see Fig. 1a) from 2009 Oct. 23 (MJD=55127) to 2010 Jun. 22 (MJD=55369). The evolution of the hardness-ratio 
(HR=BAT/GSC count rate) is shown in Fig. 1b. 

We also extracted the light curves from PCU2 data of RXTE PCA instrument to study the X-ray intensity variation of the source XTE J1752-223 during its 
2009-10 outburst (see Fig. 4a). Depending on the nature of the variation daily average count rates or outburst profiles, BHCs are mainly divided into 
two types: one is the `$fast~rise~slow~decay$' (FRSD) type and the other one is the `$slow~rise~slow~decay$' (SRSD) type (Debnath et al. 2010). Though 
there is the absence of data for one to two months in all three satellite instruments due to the sun constraint, we may classify 2009-10 outburst of 
XTE~J1752-223 as `SRSD' type based on profile light curves shown in Fig. 1a and Fig. 4a. 

Both BAT and GSC count rates rapidly increased during 2009 Oct. 23-26 (MJD=55127-55130), when PCA was not observing the source. After that count rates 
increased slowly for the next few months. Fig. 1a shows that the high energy BAT data reaches its peak flux much before the low energy GSC data. BAT 
count rate also starts to fall roughly one month before GSC. It is also evident that except during the middle phase of the outburst, there is dominance 
of the hard (BAT) flux over the soft (GSC) flux. 

\subsubsection{Evolution of Dynamic Photon Index ($\Theta$)}

During the non-operational period of RXTE PCA from 2009 Nov. 16 to 2010 Jan. 18, no data were available for spectral analysis. To study the 
spectral properties and check the existence of soft state, we calculated the dynamic photon index ($\Theta$) using the method mentioned in 
Ghosh \& Chakrabarti (2019). The RXTE ASM operates in $1.5-12$ keV energy range and has an detector area of $90~ cm^2$. There are three ASM energy 
bands (A, B, C). The band energy ranges are $1.5-3$ keV, $3-5$ keV and $5-12$ keV with average energy $E_A=2.25$ keV, $E_B=4$ keV, and $E_C=8.5$ keV 
respectively. If a, b, c represent photon count rates in A, B, C bands, then $\Theta$ is given by, $$\Theta = tan^{-1}[\frac{(c-b)}{(E_C-E_B)}]. \eqno (1)$$ 
$\Theta$ defines the slope of the hard region of the spectrum. According to Ghosh \& Chakrabarti (2019), in soft states $\Theta \to -1.57$ while 
in hard states $\Theta \to 0$. But, since the C band has a broader energy range, there is a possibility of existence of $c>b$, which gives 
$\Theta > 0$ indicating a harder state. In Fig. 3(a), we have shown the variation of $\Theta$ during the present outburst of XTE~J1752-223. 
In Fig. 3(b), variation of intensity with $\Theta$ is shown. From these variations, we can say that the source was in HS at the beginning of the outburst 
and then made transition to softer (SIMS or SS) states via HIMS and then again went back to HS. After MJD $\simeq$ 55217.8, $\Theta$ reached to a value 
$\leq -1$ and the intensity was also high around this time (Fig. 3(b)). This allows us to infer that during this higher `-'ve $\Theta$ period of the 
outburst, the source could have been in the SS.

\subsubsection{Evolution of Hardness Ratio}

Although to understand accretion flow dynamics and spectral properties of the black holes one needs to do both spectral and timing analysis, to 
understand the overall nature of the outburst, one can study the hardness ratio (HR) variation during the outburst. Here, we define the ratio 
between daily averaged Swift BAT fluxes in $15-50$~keV with MAXI GSC fluxes in $2-10$~keV as HR. 

Higher HR values ($\sim 3-4$) are observed during initial $\sim 80$ days of the outburst till 2010 Jan. 12 (MJD=55208.64). After that, it 
decreases rapidly to $\sim 0.1$ within the next $\sim 10$~days (see, Fig. 1b). If one looks at Fig. 1(a), starting from MJD=55208.64, BAT flux 
is found to decrease rapidly till it attained its lowest value (26.84 mCrab) on MJD=55222.33. Note that there was no BAT observation in between 
MJD=55218.18-55221.35. So, the lowest BAT count may have been attained earlier. A reverse nature is observed in the GSC flux. A rapid increase 
in GSC rate is observed starting from 2010 Jan. 12 (MJD=55208.64) and it attained a maximum flux on 2010 Jan. 22 (MJD=55218.80). Since on this 
GSC peak flux day, the BAT data was not available, we are unable to calculate HR. Seeing the evolution of BAT, GSC fluxes and HRs, we may define 
MJD=55208.64 as transition day between HS (rising) to HIMS (rising) and MJD=55218.80 as transition day between HIMS (rising) 
to SIMS or SS,  i.e., phase between first observation 2009 Oct. 23 (MJD=55127) to 2010 Jan. 12 (MJD=55208.64) as rising hard state, from 2010 
Jan. 12 to 2010 Jan. 22 (MJD=55208.64-55218.80) as rising HIMS. To confirm these statements, we made detailed spectral as well as temporal (nature 
of QPOs, if present) study of the source. 

The low HR was observed for the next $\sim 2$~months, till 2010 Mar. 24 (MJD=55279.63). After that it started to rise rapidly due to rise in BAT 
flux and fall in GSC flux. This phase continued for the next $\sim 10$~days, till 2010 Apr. 03 (MJD=55289.36). After that HR decreases steadily up 
to the end of the observation on the 2010 Jun. 22 (MJD=55368.94). From the nature of the variation of HR, we may define the period Jan. 23 to Mar. 
24 (MJD=55218.80-55279.63) of 2010 as the SIMS or SS, 2010 Mar. 24 to Apr. 03 (MJD=55279.63-55289.36) as the declining HIMS and the rest of the 
observations as declining hard state. So, we may say MJD=55279.63 as the transition day from SIMS or SS to HIMS (declining) and MJD=55289.36 
as that of HIMS (declining) to HS (declining). The later transition is also significant, since on this day BAT flux reaches local 
maximum during the declining phase. However, detailed spectral analysis and study of the nature of QPOs (if present) are required to confirm these 
transition dates.

To get a clearer picture of the evolution of HR, we plot the hardness intensity diagram (HID) in Fig. 2, which shows the variation of HR 
with combined GSC (2-10~keV) plus BAT (15-50~keV) flux (in mCarb) in Fig. 2. The points A, B, C, D, E, F mark start/end/transition between spectral 
states.
 
\subsubsection{QPO Evolution}
 
We carry out the fast Fourier transformed (FFT) power density spectra (PDS) using 0.01~sec binned RXTE PCU2 light curves and search for the low 
frequency QPOs. We find prominent signatures of QPOs only in $6$ observations out of the $32$ observations which we study. All the observations 
(MJDs = 55215.93, 55216.98, 55217.90, 55218.16, 55218.82, 55220.71) are found in HIMS (rising) and SIMS as stated earlier. The detailed 
nature of the dominating primary QPOs, based on Lorentzian model fits in {HEASoft's XRONOS package are given in Table 1. A monotonic rise 
in type-C QPO frequency, in $2.20-6.15$~Hz from 2010 Jan. 19 to 22 (MJD=55215.96-55218.16) is observed during HIMS (rising) and rest two 
type-B QPOs ($3.51$~Hz and $2.19$~Hz) are observed sporadically in SIMS. This nature of monotonic evolution of QPO frequency during HS and HIMS 
spectral states and sporadic QPOs during SIMS are quite common in outbursting BHCs (for more details see, Debnath et al. 2013 and references 
therein). But strangely we have not seen any QPOs during the rising hard and declining HS and HIMS spectral states. It is possible that the 
resonance condition between the Compton cooling and compressional heating did not take place as is considered to be the cause of low-frequency 
QPOs in such cases (Molteni et al. 1996; Chakrabarti et al. 2015).

\subsection{Spectral properties}    

To study the spectral properties of the BHC XTE J1752-223 during its 2009-10 outburst, we use archival data of RXTE PCU2 instrument from October 
26, 2009 (MJD 55130.96) to June 21, 2010 (MJD 55368.94) in the energy band of $2.5-25$~keV. To find rough estimation of the degree of importance 
of the thermal and the non-thermal components in the emitted spectra, we first fitted $32$ spectra with the combined DBB and PL models. Then to 
understand physical picture of the accretion flow dynamics during the outburst, we then refitted all spectra with the TCAF model {\it fits} file 
as a local additive table model in XSPEC. In Table 2, we mention spectral parameters obtained from the spectral fits from two different types of 
model fits. To see the overall nature of the outburst, in Fig. 4 we show compare the variations of the PCU2 count rate (in 2-25 keV), total flux 
(DBB+PL) and total rate (${\dot{m}}_d$ + ${\dot{m}}_h$). The evolution of DBB and PL model component fluxes and parameters (DBB inner-disk 
temperature T$_{in}$ in keV and PL photon index $\Gamma$) are shown in Fig. 5. In Fig. 6, we show the variation of the TCAF model fitted spectral 
parameters. We also make the HID and ARRID and show them in Fig. 2 and Fig. 7 respectively. Looking at the HID and ARRID, we notice that both of 
them show a hysteresis behavior. Six points (A, B, C, D, E, F) are marked in HID and ARRID (here in the ARRID we could not specify the location 
of point B due to absence of data) to define spectral state transitions. B, C, D, E points represent the transition days from HS (rising) $\to$ 
HIMS (rising), HIMS (rising) $\to$ SIMS or SS, SIMS or SS $\to$ HIMS (declining) and HIMS (declining) $\to$ HS (declining) respectively.

Depending upon the nature of variation of TCAF model fitted flow parameters, PL photon index, the nature of QPOs, HID and ARRID, we classified 
the entire 2009-10 outburst of XTE~J1753-223 into the following spectral states (HS, HIMS and SIMS or SS), observed in the following sequence: HS 
(rising) $\rightarrow$ HIMS (rising) $\rightarrow$ SIMS or SS $\rightarrow$ HIMS (declining) $\rightarrow$ HS (declining). 
This classification also supports our rough understanding of the spectral nature of the source based on the evolution of the temporal properties, 
i.e., based on variation of GSC, BAT fluxes and HR. Our spectral analysis confirms middle phase of the outburst as SIMS or SS.

\noindent{\it $i)$ Hard State -- Rising Phase : }
From the spectral analysis, we may define initial period of the outburst, prior to 2010 Jan. 19 (MJD=55215.91) as HS due to high dominance of the 
sub-Keplerian halo rate (${\dot{m}}_h$) over the Keplerian disk rate (${\dot{m}}_d$), presence of strong shock located far away from the BH and 
low value of the PL index ($\Gamma \sim 1.4$). ARR was also found to have a higher value. Note that RXTE PCA was not operational during 2009 Nov. 
16 (MJD=55151) to 2010 Jan. 18 (MJD=55214). So, to know  the exact transition date of HS (rising) to HIMS (rising), we needed to rely 
on the variation of HR (see, Fig. 1b), and found that it actually occurred on 2010 Jan. 12 (MJD=55208.64). HR started to fall rapidly due to rapid 
rise in soft band (GSC) flux and rapid decrease in hard band (BAT) flux. This is marked as B in HID in Fig. 2. No prominent signature of QPOs are 
observed during this phase of the outburst, which may be due to non-satisfaction of the resonance condition to form oscillating shock as mentioned 
earlier.

\noindent{\it $ii)$ Hard-Intermediate State -- Rising Phase : }
On the first observation day (MJD=55215.91), the source was already in HIMS (rising). The combined spectral and temporal analysis suggest 
that it continued till 2010 Jan. 22 (MJD=55218.80). During this phase of the outburst, disk rate observed is found to rise slowly. But halo rate 
is observed at higher values. We also see evolving low frequency type-C QPOs (from $2.20$~Hz to $6.15$~Hz) during this phase. In this phase the 
hardness falls rapidly in HID (Fig. 2) and ARR also went to a lower value in ARRID (Fig. 7). There are two observations on MJD=55218. In the first 
observation (MJD=55218.14), we see a QPO of maximum observed frequency. In the other observation (MJD=55218.80), a type-B QPO of much lower frequency 
(=$3.51$~Hz) is observed. On this later observation we see dominance of the disk rate over halo rate. So, we define this observation as HIMS 
(rising) to SIMS transition day.

\noindent{\it $iii)$ Soft-Intermediate/Soft State : }

We assign the spectral state from 2010 Jan. 22 (MJD=55218.80) to 2010 Mar. 24 (MJD=55279.63) as the SIMS or SS. Here, we observe characteristic sporadic 
type-B QPOs of frequencies $3.51$~Hz and $2.19$~Hz. During this phase of the outburst, ${\dot{m}}_d$ dominates over ${\dot{m}}_h$ in the presence of 
a weaker shock. Although as the day progresses, supply in the Keplerian disk decreases slowly. PL index $\Gamma$ is observed in between $2.1-2.2$. 
Low HR due to dominating GSC flux over BAT flux and low ARR due to dominating Keplerian disk rate over sub-Keplerian halo rate is also observed during 
this phase of the outburst. This also supports our spectral classification. We identify MJD=55289.36 as the transition day from SIMS or SS to HIMS 
(declining) since after this observation the flow rate as well as PL flux started to increase. 

\noindent{\it $iv)$ Hard-Intermediate State -- Declining Phase : }
The rapidly evolving period from 2010 Mar. 24 (MJD=55279.63) to 2010 Apr. 03 (MJD=55289.36) is assigned to be in  HIMS in the declining phase. During 
this phase, the shock was observed to move away quickly from $35$ to $126~r_s$. Decrease in DBB flux and increase in PL flux are observed. HR is found 
to increase rapidly. ARR also increased in this period. On the last day, i.e., the day of transition from the HIMS (declining) to HS 
(declining), much higher halo rate is observed as compared to the low disk rate when the ARR jumped to a very higher value. On this day 
(MJD=55289.36), a strong shock (R=$2.6$) is also observed. We have not seen any low frequency QPO during this phase of the outburst.

\noindent{\it $v)$ Hard State -- Declining Phase : }
The source was observed in this spectral state starting from HIMS (declining) to HS (declining) transition day (2010 Apr. 03; 
MJD=55289.36), till the end of the observation (2010 Jun. 06; MJD=55368). Low value of the PL indices ($\Gamma \sim 1.47-1.75$) are observed 
during this phase of the outburst. Both DBB and PL fluxes are found to decrease. Although dominance of ${\dot{m}}_h$ is observed it decreases 
as the day progresses. During this phase, presence of strong shocks located far away from the black hole is observed. 

\subsection{Accretion Rate Ratio Intensity Diagram (ARRID)}
Jana et al. (2016) showed that the accretion rate ratio intensity diagram (ARRID) gives more physical picture than hardness-intensity 
diagram (HID) during their study of the 2011 outburst of MAXI~J1836-194. In Fig. 2, HIDs using photon fluxes of MAXI GSC (2-10~keV) and Swift 
BAT (15-50~keV) are plotted. Here, to get a better understanding of the correlation between spectral and temporal properties and spectral state 
transitions from physical point of view based on TCAF model fitted disk and halo rates, we have plotted ARRID in Fig. 7. In ARRID, PCA count 
rates are plotted as a function of ARR ($\dot{m_h}/\dot{m_d}$). In Fig. 7, we have shown the variation of $2-25$~keV PCU2 rate with ARR. Spectral 
state transition dates (points C, D, E with MJDs are 55218.80, 55279.63 and 55289.36 respectively) are more clear than what is seen from HID. Due 
to the non-operational of RXTE PCA, we are unable to pinpoint B when the transition from HS (rising) to HIMS (rising) was occurred. During initial 
phase of the outburst X-ray intensity increases with roughly constant ARR in the HS (rising). Then we see that the ARR decreases rapidly with 
increasing intensity. This indicates that the source is making transition into intermediate states. Intensity starts to fall rapidly after point 
C, where a transition from HIMS (rising) to SIMS or SS is occurred. During this phase, ARR also decreased marginally. Point D implies transition from 
SIMS or SS to HIMS (declining), as after this point ARR starts to slowly increase first (with rise in intensity), then rapidly. A sharp fall in count 
rate with roughly constant ARR is observed after point E, where a transition from HIMS (declining) to HS (declining) was observed.

\subsection{Mass Estimation from Spectral Analysis}
Since in TCAF {\it fits} file, mass of the black hole is an important model input parameter,  while fitting spectrum with it, we may keep it free 
if it is not known. Here, we keep mass of XTE~J1752-223 as free and obtain its variation in between $8.1-11.9$~$M_{\odot}$. This variation of mass 
of the black hole is not the actual variation during our analysis period, rather it is a fluctuation due to errors in measurements. This may occur 
due to errors contributed from various factors, for example instrumental errors, which may not be the same for all observations. Also, mass of the 
black hole is highly sensitive to the temperature of the disk ($M_{BH} \sim T^4$), i.e., any small error in determination of temperature gives rise 
to a significant error in the fitted mass value. Keeping in mind of all of these measurement errors, from our spectral analysis, we predict the 
probable mass of XTE~J1752-223 as $8.1-11.9$~$M_\odot$, or $10\pm{1.9}~M_\odot$. Here, $10~M_\odot$ is the average value of TCAF model fitted masses.

\section{Discussions and Conclusions}

Galactic transient BHC XTE~J1752-223 was monitored in multi-waveband during its 2009-10 outburst immediately after discovery by RXTE all sky monitor 
(ASM) on 2009 Oct. 21 (MJD=55125). RXTE PCA monitored this source starting from 2009 Oct. 26 (MJD=55130.93) roughly on a daily basis except 
from 2009 Nov. 16 to 2010 Jan. 18 (operational shutdown period due to Sun constraint). We found that non-operational period of MAXI and Swift satellites 
were less as compared to RXTE. So, to study outburst profiles (daily average count rates) and hardness-ratio (HR) we use MAXI GSC (in $2-10$~keV) and 
Swift BAT (in $15-50$~keV) data from 2009 Oct. 23 (MJD=55127) to 2010 Jun. 22 (MJD=55369). We also made use of the RXTE ASM data (daily average light 
curve) to make dynamic photon index ($\Theta$), to check the existence of soft state during the outburst. Spectral analysis is made using RXTE PCU2 data 
(in $2.5-25$~keV). Here we analyzed $32$ observations spreaded over the outburst period. Two types of models such as $i)$ commonly used combined DBB plus 
PL model, and $ii)$ physical accretion flow model namely TCAF model, are used for our spectral analysis. 

Although low frequency QPOs are very common in this type of stellar mass BHCs, we observed them only in six observations out of the  $32$ observations 
under consideration. Evolving type-C QPOs of frequencies from $2.20$ to $6.15$~Hz are observed during 2010 Jan. 19-21. Later we observed sporadic type-B 
QPOs. All these QPOs are observed during the rising phase of the outburst. Depending upon the nature of evolution of spectral (model fitted parameters) 
and temporal ($\Theta$, HR, QPO) properties, we make an effort to understand spectral nature of the source during the outburst. We find the presence of 
these major spectral states: HS, HIMS and SIMS or SS. The outburst started and ended with HS via intermediate states to form a hysteresis loop in the 
following sequence : HS (rising) $\rightarrow$ HIMS (rising) $\rightarrow$ SIMS or SS $\rightarrow$ HIMS (declining) $\rightarrow$ HS 
(declining).  

We classified the spectral state during 2009 Oct. 23 (MJD=55127) to 2010 Jan. 12 (MJD=55208.64) as the rising HS. During this phase of the outburst, high 
values of HR ($\sim 3-4$) as well as PL flux or halo rate are observed. From our spectral study of initial four spectra, we see highly dominating halo 
rate in presence of strong shock located far away from the BH. High DBB temperature ($T_{in} \sim 1.25-1.44$~keV) and low PL index ($\Gamma \sim 1.35-1.39$) 
values are also observed. Although, we have not been able to make spectral analysis on MJD=55208.64, based on variation of HR, we identify this as HS 
(rising) to HIMS (rising) transition day. After this observation, HR decreases rapidly due to rapid fall in BAT count and rise in GSC count. 
This phase (rising HIMS) is continued for the next $\sim 10$~days till 2010 Jan. 22 (MJD-55218.80), when we observe HR as $0.1$. Our spectral and temporal 
analysis also support this. Dominating halo was found to form a strong shock. During this phase of our outburst, we also see monotonically evolving low 
frequency QPOs ($2.20-6.15$~Hz). These evolving QPOs are characteristic feature of HS and HIMS (see, Debnath et al. 2013 and references therein). We define 
MJD=55218.80 as HIMS (rising) to softer state transition day as on this day, we see that GSC flux reaches at its maximum. On this day, there are two 
PCA observations: In the first one (MJD=55218.16), we see a maximum frequency ($6.15$~Hz) of the evolving type-C QPO and on the second one (MJD=55218.82), 
we see a much lower ($3.51$~Hz) type-B QPO. On the next day, QPOs are absent and re-appeared two days later (on MJD=55220.71 at $2.19$~Hz). This sporadic
nature of the QPO is the characteristic feature of SIMS. From spectral analysis, we also see cooler disk (DBB $T_{in} \sim 0.37-0.92$~keV) with higher
 $\Gamma$ ($\sim 1.84-2.22$). Decreasing PL flux or halo rate with higher DBB flux or disk rate are observed during this phase of the outburst. 

We identify 2010 Mar. 24 (MJD=55279.63) as SIMS or SS to HIMS (declining) phase. This is because, after this phase, we see rise in HR due to faster 
increase in BAT count. This increase in HR is continued for the next $\sim 10$~days till 2010 Apr. 03 (MJD=55289.36). Our spectral analysis also shows 
that during this phase of the outburst, increasing trend of $T_{in}$ with fall in DBB flux and rise in PL flux. PL index $\Gamma$ also shows a decreasing
nature. Decreasing disk rate in presence of receding shock is observed during this phase. MJD=55289.36 is identified as the transition day between HIMS 
(declining) to HS (declining) as on this day, we see higher HR due to local maximum of BAT count. After that both BAT and GSC counts, DBB 
and PL fluxes and disk and halo rates decrease. Noticeably, on the transition day we see much lower disk rate as compared to highly dominating halo 
rate in presence of a receding strong shock. Interestingly, we have not seen any signature of evolving QPOs during declining HS, HIMS as well as rising 
HS. Since these QPOs (generally type-C) form due to resonance between compressional heating and Compton cooling time scales inside CENBOL (Molteni 
et al. 1996; Chakrabarti et al. 2015), it is possible that the timescales did not match as the halo rate was too large.

It is well known that the spectral and the temporal properties are strongly correlated. HID or so-called `q'-diagram (see, Fig. 2) gives 
a quick-look of it, where different spectral states are linked with the different branches of the diagram. In Fig. 2, A-B, E-F mark HS of the rising and 
the declining phases respectively, where as B-C, D-E indicate HIMS of the rising and the declining phases respectively. Here, C-D marks SIMS or SS. 
In 2016, Jana et al. reported that ARRID (see, Fig. 7) provides more physical picture than HID. In Fig. 7, one could find more clear indication or turn-over 
in the plot nature when a spectral state transition occurs (see for example, points C, D, E).

This source was extensively studied during its only outburst (2009-10) till date by many researchers to explore its properties in multi-waveband. 
Shaposhnikov et al. (2010) studied both spectral and timing properties of the source using RXTE PCA data and classified entire outburst 
into low-hard (LHS), intermediate (IS) and high soft (HSS) spectral states. Nakahira et al. (2010) and Curran et al. (2011) also studied independently 
with data of MAXI and Swift satellites and found similar spectral states including HSS. These reports of presence of SS or HSS, during 
2009-10 outburst of XTE~J1752-223 is quite unusual for the object as the source belongs to a class of objects of short orbital period which generally
 do not show SS during an outburst. We therefore were motivated to verify these reported states using the physical TCAF fits. Fig. 1(c), suggests 
that the source was highly active in radio during entire phase of the outburst (also see, Brocksopp et al. 2013). It is quite unusual to see jets 
during SS, if we consider radio has been emitted from jets. So, to confirm if the source actually went in SS or not, we calculated dynamic photon index 
($\Theta$) using RXTE ASM data as method described in Ghosh \& Chakrabarti (2019). $\Theta$ is found to be $< -1$ during MJD=55219 to 55233, which 
indicates the presence of a softer state, although during the entire outburst, the power-law photon index ($\Gamma$) never crossed a value $>2.3$. 
So, we term this whole period (from MJD 55218.80 to MJD 55279.63) as SIMS or SS. 

We have also estimated the disk (soft) flux percentage in $2.5-20$ keV energy band from DBB+PL model fitted spectra and power continuum (r), integrated 
over $0.1-10$ Hz in PDS. These analysis were done using methods defined in Remillard (2004). The disk i.e., DBB flux percentage was found more than $75 \%$ 
during the period from MJD=55221.3 to MJD=55279.6 (see, Table 3). Now, according to Remillard (2004), this period of observation should be defined as SS. 
However, this paper also suggests that if $r$ is $< 0.06$, that could also be the indication of the presence of SS. Here, $r$ was less than $0.06$ from 
the day when RXTE rediscovered the source (on MJD 55215.9) after Sun constraints till the end of our studied period. So according to him, this day should be 
the beginning of the SS as there was no prior PCA observation. However, there were presence of type-C QPOs on four days (MJD 55215.9, 55216.9, 55217.8 and 55218.1) 
and type-B sporadic QPOs on two days (MJD 55218.8 and 55220.6) on which $r$ was $< 0.06$. The QPO information and spectral fitted parameters tell us 
that the source was still in intermediate state (HIMS and SIMS) at-least until MJD=55220.6. The dynamic photon index ($\Theta$ as defined earlier) was 
also found to be less than $-1$ from MJD=55218.8 to 55233.0, which again tells us that the softer states (SIMS or SS) could be started on MJD=55218.8. 
For the above contradictions, we found difficulty to define exact transition day between SIMS and SS and vise-verse and designated whole middle phase 
(from MJD 55218.8 to MJD 55279.6) of the outburst as mixture of two softer spectral states i.e., as `SIMS or SS'.

Optically thin radio flares were also present during this time period of the outburst (see Fig. 1(c)). During accretion, matter can bring in a large 
amount of stochastic magnetic flux tubes which due to azimuthal velocity, form a very strong toroidal flux tubes (Nandi \& Chakrabarti 2001). The 
magnetic tension acting on these flux tubes could be very strong collapsing toroidal flux tubes and forming sporadic outflows. We believe that the 
disk could be magnetically dominated in order to see these jets. We calculated equipartition magnetic field strengths during the flaring as well as 
intermediate phases and found presence of strong magnetic field in the region from MJD=55215 to 55285. We believe that this high magnetic field acts 
as a driving force for the optically thin radio flares in the SIMS or SS. The details calculation of magnetic field and its effect and 
correlation with radio flares are beyond the scope of this paper and will be published elsewhere.

Recent studies using TCAF confirmed masses of many BHCs from the spectral analysis (Molla et al. 2016, 2017; Jana et al. 2016, 2019; Chatterjee et 
al. 2016, 2019; Debnath et al. 2017). In TCAF, mass of the BH is an important model input parameter. So, by fitting (keeping it as free while fitting) 
one gets the best fitted masses. Here we see the variation of the model fitted masses between $8.1-11.9$~$M_{\odot}$. The average observed mass is 
$10$~$M_{\odot}$. So, we believe that the probable mass range of the source is $8.1-11.9$~$M_\odot$, or $10\pm{1.9}~M_\odot$. This estimated mass of 
the source falls roughly in the same range as predicted by Shaposhnikov et al. (2010). They estimated mass of source as $8-11$~$M_\odot$ or 
$9.8{\pm 0.9}~M_\odot$ from their QPO frequency vs. spectral slope correlation method.

\section*{Acknowledgements}
This work made use of ASM and PCA data of NASA's RXTE satellite; Swift BAT data provided by UK Swift Science Data Centre at the 
University of Leicester; and MAXI GSC data provided by RIKEN, JAXA and the MAXI team. 
K.C. acknowledges support from DST/INSPIRE fellowship (IF170233).
D.D. and S.K.C. acknowledge support from Govt. of West Bengal, India and ISRO sponsored RESPOND project (ISRO/RES/2/418/17-18) fund.
D.C. and D.D. acknowledge support from DST/SERB sponsored Extra Mural Research project (EMR/2016/003918) fund.
A.J. and D.D. acknowledge support from DST/GITA sponsored India-Taiwan collaborative project (GITA/DST/TWN/P-76/2017) fund.
A.J. also acknowledges CSIR SRF fellowship (09/904(0012)2K18 EMR-1).


\clearpage

\begin{figure}
\centering
       \includegraphics[width=3.0in]{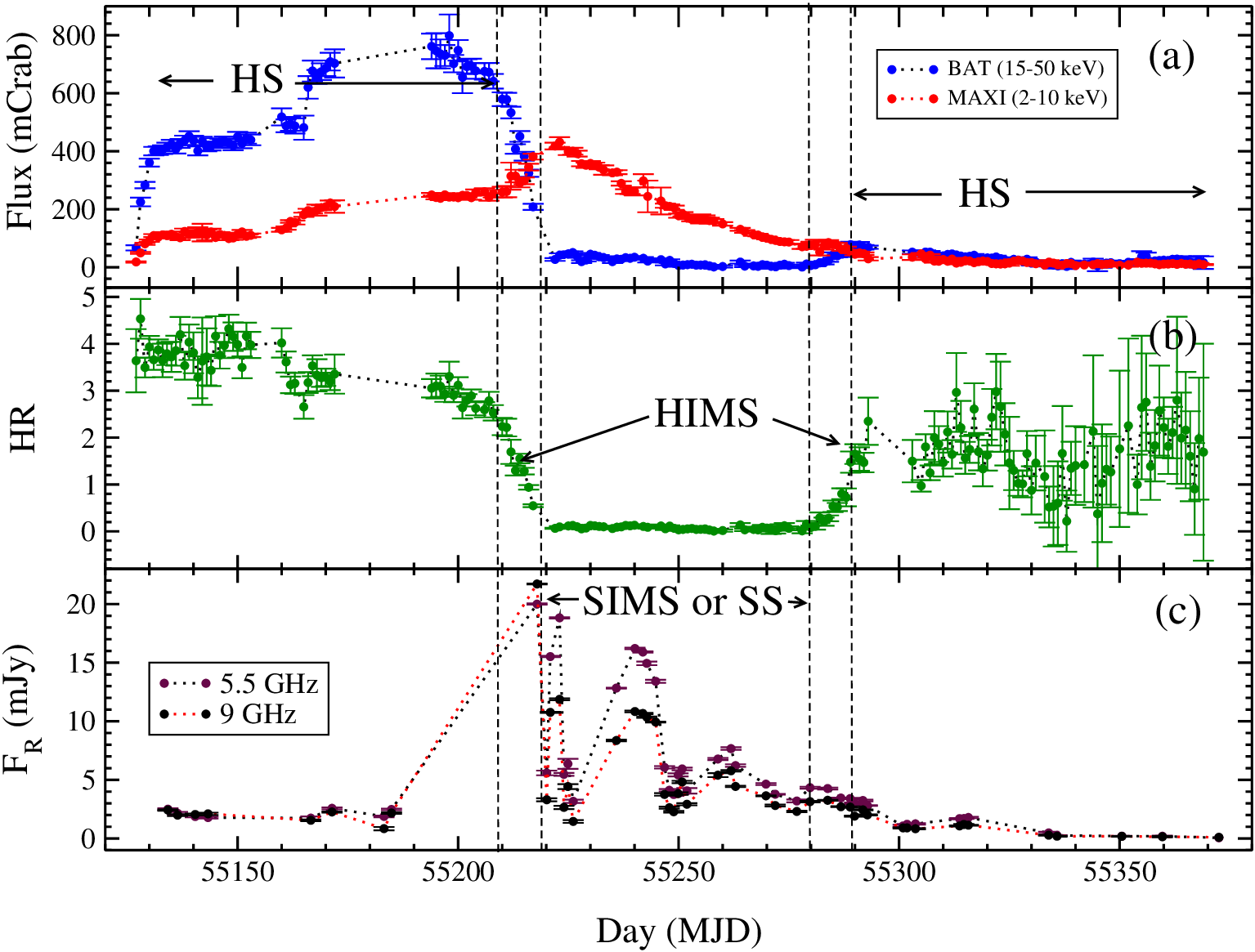}
\caption{Variation of (a) $2-10$~keV MAXI GSC (dotted-point online red curve) and $15-50$~keV Swift BAT (dashed-point online blue curve)
one day average fluxes in units of mCrab, (b) variation of the hardness-ratios (HRs=BAT/GSC fluxes) and (c) radio light curves (at 5.5 and 
9 GHz) during the 2009-10 outburst of XTE~J1752-223 are shown. The vertical dotted lines mark transition between observed spectral states.}

\label{f1}
\end{figure}

\begin{figure}
        \centerline{
                   \includegraphics[scale=0.6,width=8truecm,angle=0]{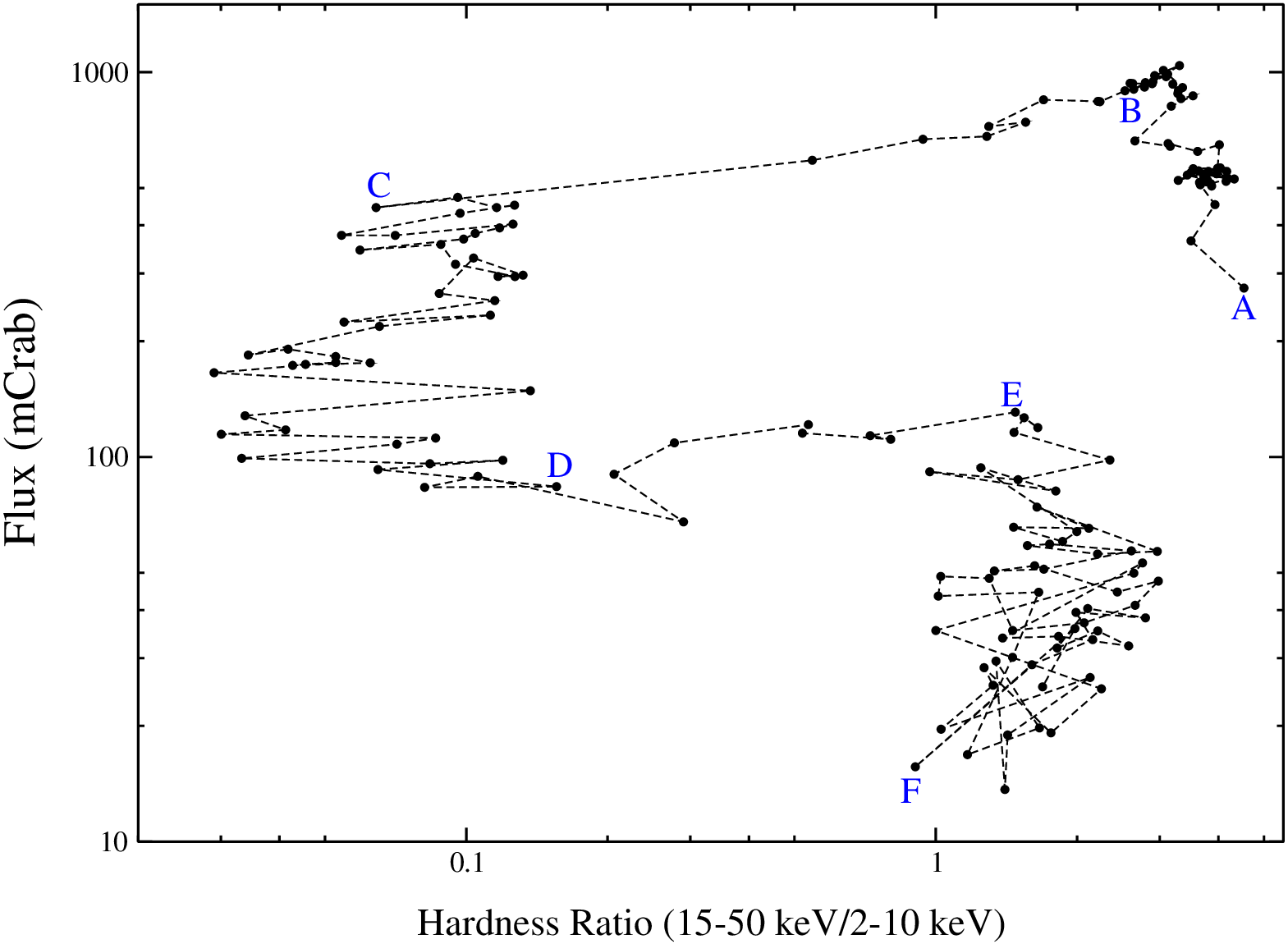}
                   }
\caption{Variation of total (2-10 keV MAXI GSC + 15-50 keV Swift BAT) flux (in mCrab) with HR (BAT/GSC fluxes) during
         the 2009-10 outburst for XTE J1752-223 is shown. Here A, B, C, D, E and F points mark the start, the end or
         the spectral state transitions. }
\label{fig2}
\end{figure}

\begin{figure}
\centering
       \includegraphics[scale=1.0,width=3.5in]{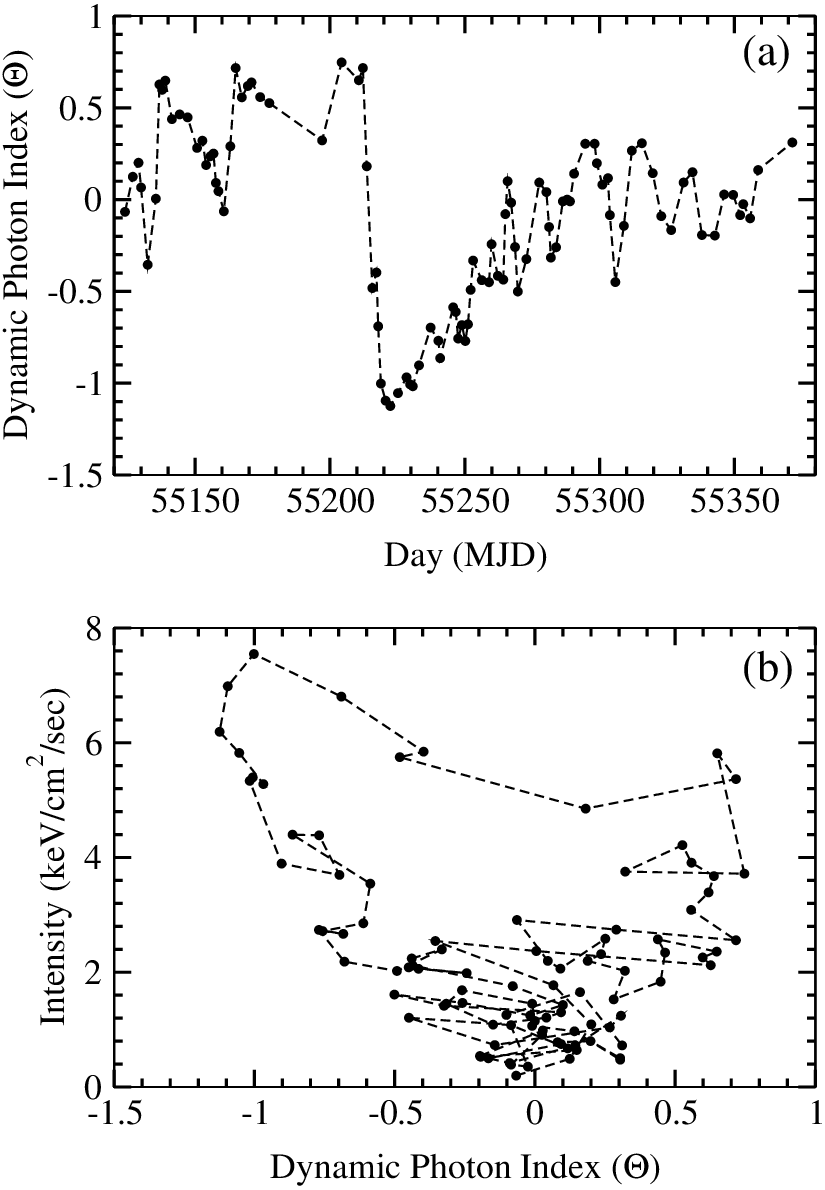}
\caption{Variation of (a) dynamic photon index ($\Theta$) using RXTE ASM data and (b) 1.5-12 keV RXTE ASM intensity
         (in $keV/cm^2/sec$) with $\Theta$ during the 2009-10 outburst of XTE J1752-223 is shown.}
\label{f3}
\end{figure}

\begin{figure}
 \vskip 0.8cm
        \centerline{
                   \includegraphics[scale=0.6,width=8truecm,angle=0]{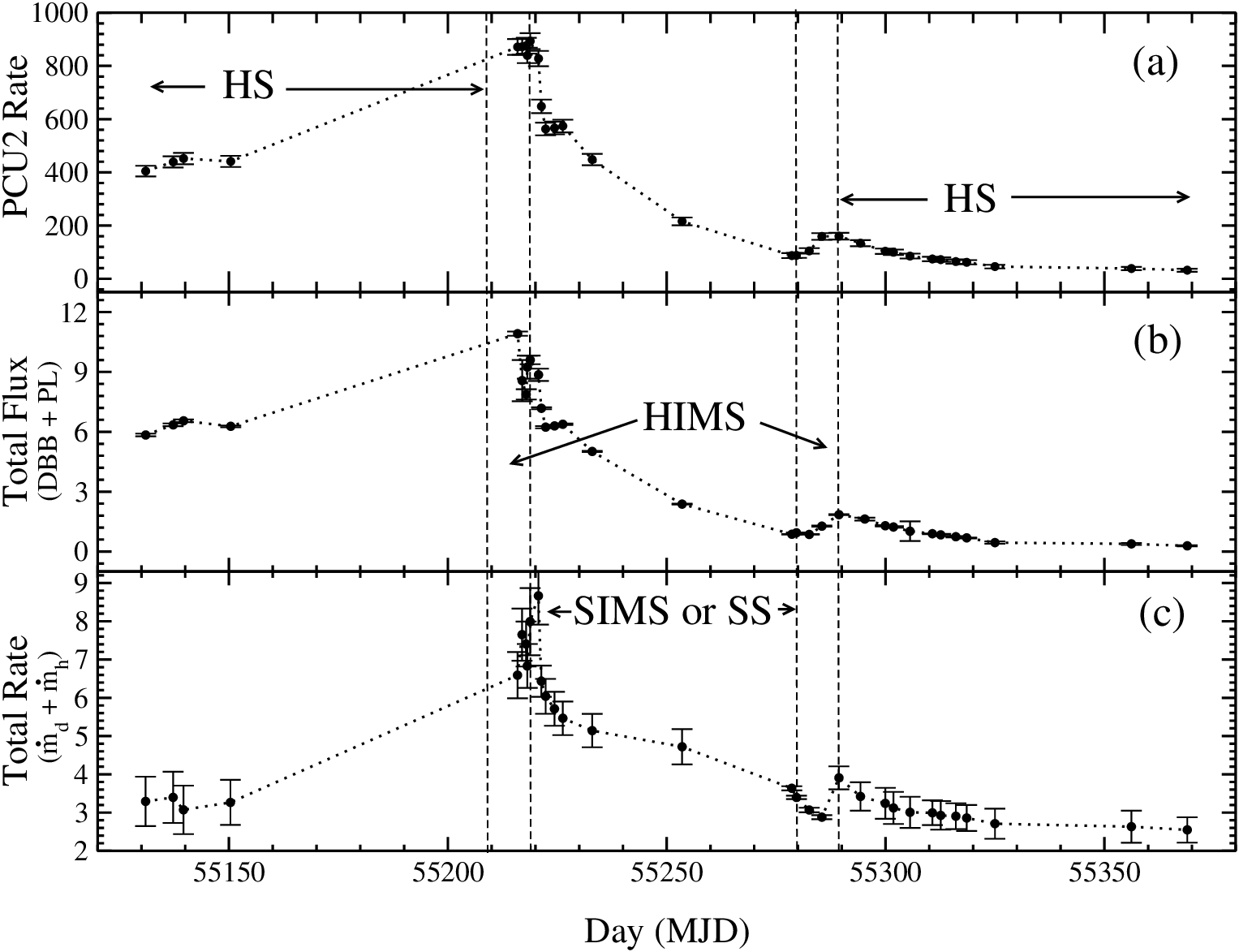}
                   }
\caption{Variations of (a) 2-25 keV PCA count rate (cnts/s), and combined DDB plus PL model fitted (b) total flux (DBB+PL flux 
in $10^{-9}~erg~cm^{-2}~s^{-1}$), TCAF model fitted (c) disk rate + halo rate (${\dot m}_d + {\dot m}_h$ 
in ${\dot M}_{Edd}$) in the $2.5-25$~keV spectra with time (Day in MJD) are shown. }
 \label{fig4}
\end{figure}

\begin{figure}
 \vskip 0.8cm
        \centerline{
                   \includegraphics[scale=0.6,width=8truecm,angle=0]{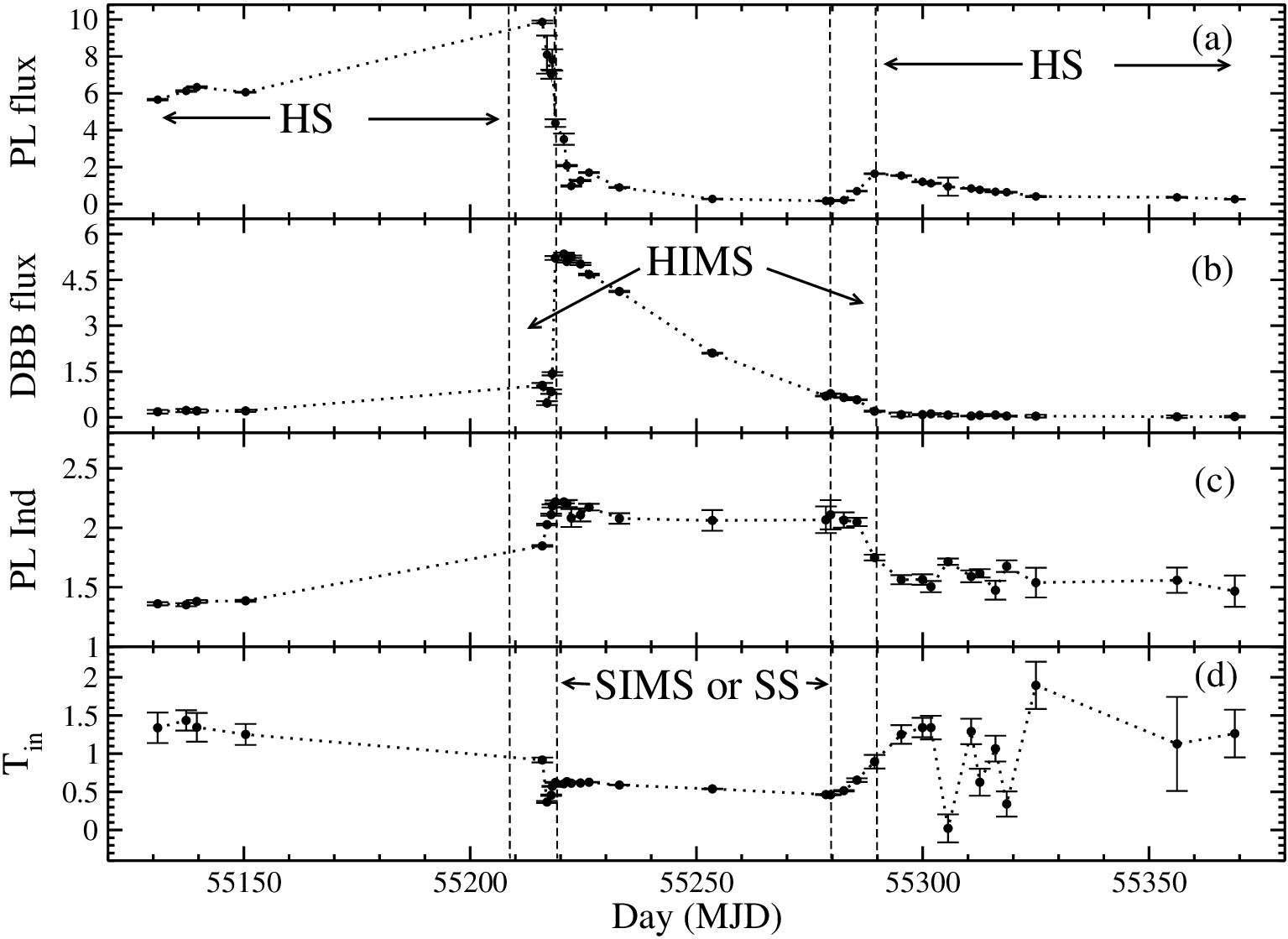}
                   }
\caption{Variation of combined DBB plus PL model fitted (a) Power-law flux (in $10^{-9}~erg~cm^{-2}~s^{-1}$), 
(b) disk black body flux (in $10^{-9}~erg~cm^{-2}~s^{-1}$), (c) photon index of power-law ($\Gamma$) and 
(d) inner-disk temperature ($T_{in}$ in keV) as a function of time (Day in MJD) are shown. }
 \label{fig5}
\end{figure}

\begin{figure}
 \vskip 0.8cm
        \centerline{
                   \includegraphics[scale=0.6,width=8truecm,angle=0]{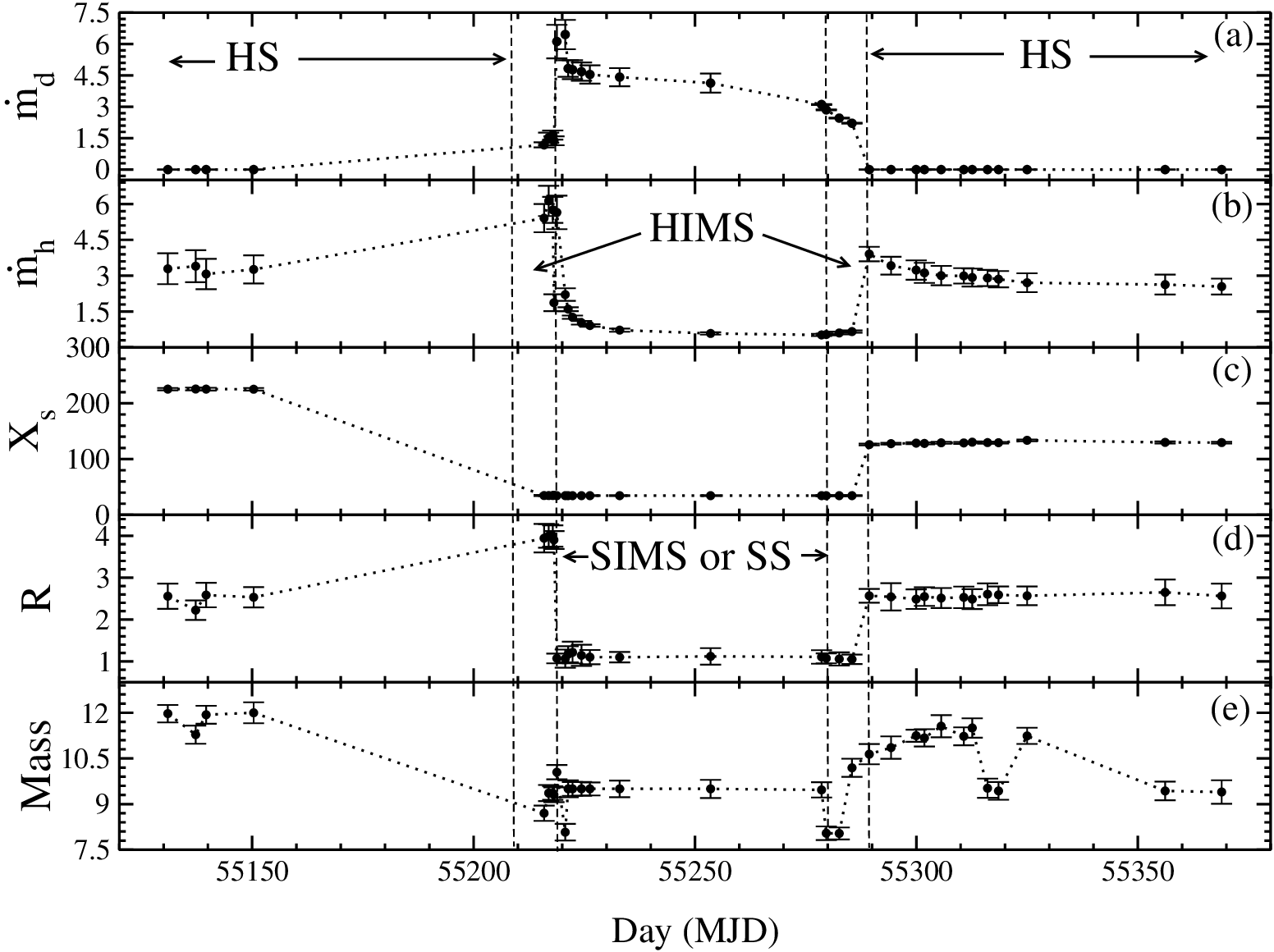}
                   }
\caption{Variations of TCAF model fitted (a) disk rate (${\dot{m}}_d$ in ${\dot M}_{Edd}$), (b) halo rate (${\dot{m}}_h$ in ${\dot M}_{Edd}$), 
(c) shock location ($X_s$ in $r_s$), (d) compression ratio ($R$), and (e) mass of the BH (in $M_{\odot}$) as a function of time 
(Day in MJD) are shown. }
 \label{fig6}
\end{figure}

\begin{figure}
 \vskip 0.8cm
        \centerline{
                   \includegraphics[scale=0.6,width=8truecm,angle=0]{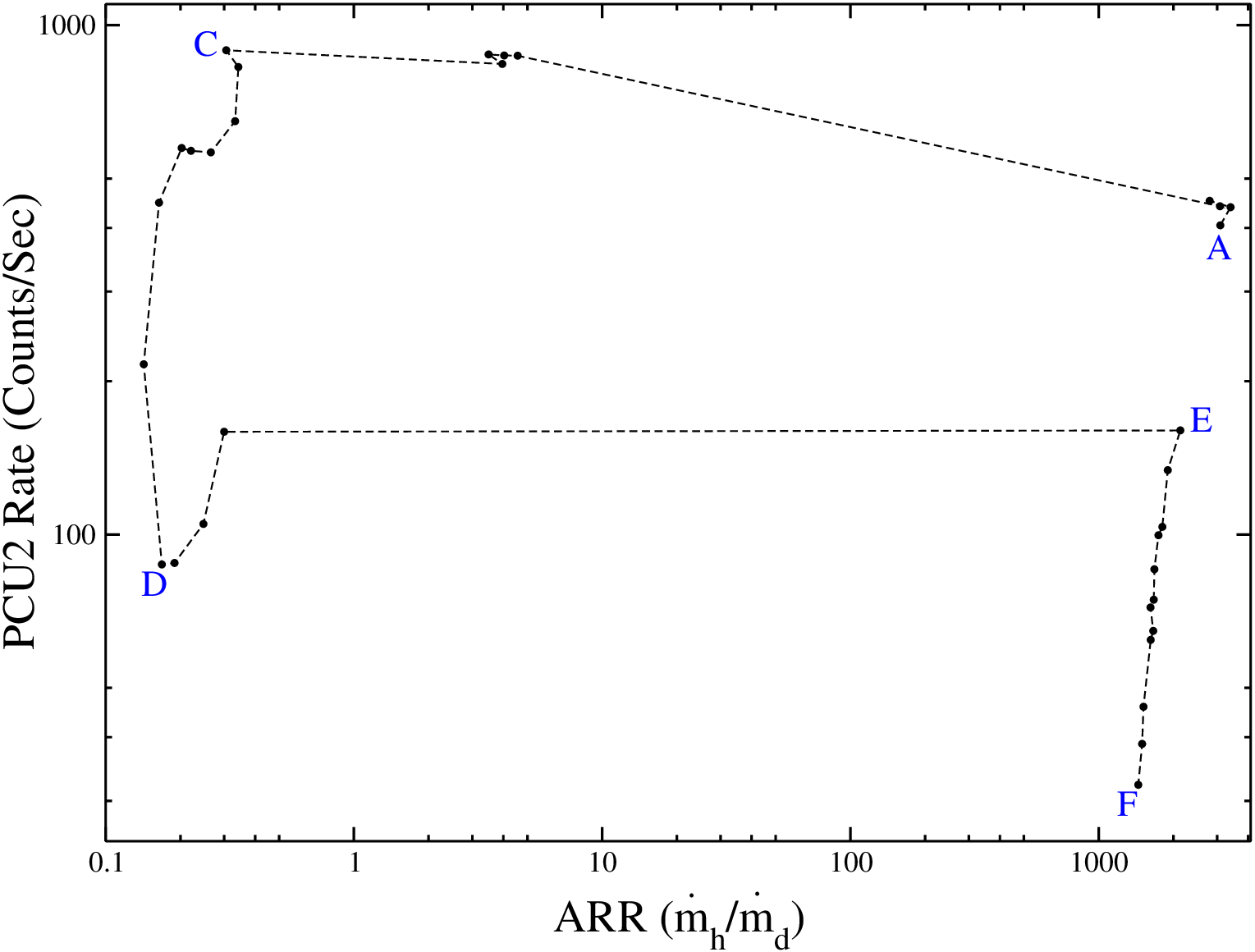}
                   }
\caption{Variation of 2-25 keV RXTE PCA PCU2 count rate with accretion rate ratio (ARR = ${\dot{m_h}}/{\dot{m_d}}$) during the 2009-10 outburst
        for XTE J1752-223 is shown. Here A, C, D, E and F points mark the start or the end or the transition between spectral states. Due to absence of    
        PCA data, here we are unable to show the desired position of B (see Fig. 2) where a transition from HS (ris.) to HIMS (ris.) was occurred. 
        Note, here X and Y axes are plotted in lagarithmic scales.}
\label{fig7}
\end{figure}

\clearpage

\begin{table*}
 \addtolength{\tabcolsep}{-2.5pt}
 \caption{QPO properties}
 \label{tab:table1}
 \begin{tabular}{lcccccc}
 \hline
  Obs Id    &     UT$^{[1]}$  &    Day     & $\nu_{qpo}$$^{[2]}$ &   $\Delta{\nu}$$^{[2]}$ & Q$^{[3]}$ &  rms (\%)\\
            &     Date        &    MJD     &      (Hz)           &   (HZ)         &           & \\
  (1)       &     (2)         &    (3)     &      (4)            &    (5)         & (6)       &   (7)\\
 \hline
  X-06-00   &  2010-01-19  &  55215.93  &  $2.20~{\pm~ 0.02}$  &  $0.70~{\pm~ 0.10}$  &  3.14  &  10.95\\
  X-06-01   &  2010-01-20  &  55216.98  &  $4.02~{\pm~ 0.03}$  &  $1.34~{\pm~ 0.18}$  &  2.99  &  10.53\\
  X-06-02   &  2010-01-21  &  55217.90  &  $5.29~{\pm~ 0.04}$  &  $1.56~{\pm~ 0.26}$  &  3.39  &  8.87 \\
  Y-01-08   &  2010-01-22  &  55218.16  &  $6.15~{\pm~ 0.16}$  &  $1.49~{\pm~ 0.54}$  &  4.10  &  4.73 \\
  Y-01-00   &  2010-01-22  &  55218.82  &  $3.51~{\pm~ 0.01}$  &  $0.69~{\pm~ 0.04}$  &  5.08  &  7.15 \\
  Y-01-02   &  2010-01-24  &  55220.71  &  $2.19~{\pm~ 0.06}$  &  $1.45~{\pm~ 0.26}$  &  1.51  &  4.50 \\
\hline
 \end{tabular}

 \noindent{
 \leftline{$^{[1]}$ `X'= 94331-01 and 'Y'= 95360-01 mark initial part of the obervation IDs.}
 \leftline{$^{[2]}$ $\nu_{qpo}$ and ${\Delta{\nu}}$ represent observed QPO frequency and its full width at} 
 \leftline{half maximum (FWHM), obtained by fitting PDS with Lorentzian profiles.}
 \leftline{$^{[3]}$ Q (=$\nu_{qpo}$/${\Delta{\nu}}$) is the coherence factor, indicates sharpness of the QPO.} 
          }
\end{table*}

\begin{table*}
\small
 \addtolength{\tabcolsep}{-4.5pt}
 \centering
 \caption{Properties of spectral model fitted parameters}
 \label{tab:table3}
 \begin{tabular}{lccccccc|ccccccc}
 \hline
  Obs ID$^{[1]}$ & UT$^{[2]}$ & MJD & DBB Fl$^{[3]}$ & PL Fl$^{[3]}$ & T$_{in}$$^{[3]}$& $\Gamma$$^{[3]}$ & ${\chi}^2$/dof$^{[5]}$ & ${\dot m}_d$$^{[4]}$ & ${\dot m}_h$$^{[4]}$ & $X_s$$^{[4]}$ & $R$$^{[4]}$ & $M_{BH}$$^{[4]}$ & ${\chi}^2$/dof$^{[5]}$ \\
  (1) & (2) & (3) & (4) & (5) & (6) & (7) & (8) & (9) & (10) & (11) & (12) & (13) & (14) \\
 \hline

 A-01-00 & 10/26& 55130.93& $0.18~{\pm ~0.05}$& $5.65~{\pm ~0.04}$& $1.34~{\pm~ 0.20}$& $1.36~{\pm~ 0.01}$& 49/43 &$0.001~{\pm~2.6E-04  }$& $3.29~{\pm~0.64 }$& $225.3~{\pm~2.1 }$& $2.66~ {\pm~0.30 }$& $11.9 ~{\pm~0.3 }$& 42/41  \\
 B-02-06 & 11/02& 55137.23& $0.22~{\pm ~0.05}$& $6.12~{\pm ~0.04}$& $1.44~{\pm~ 0.13}$& $1.35~{\pm~ 0.01}$& 50/43 &$0.001~{\pm~2.3E-04  }$& $3.39~{\pm~0.66 }$& $225.6~{\pm~2.5 }$& $2.23~ {\pm~0.23 }$& $11.3 ~{\pm~0.3 }$& 49/41  \\
 B-02-10 & 11/04& 55139.58& $0.21~{\pm ~0.05}$& $6.33~{\pm ~0.04}$& $1.35~{\pm~ 0.19}$& $1.38~{\pm~ 0.01}$& 54/43 &$0.001~{\pm~1.9E-04  }$& $3.07~{\pm~0.63 }$& $225.4~{\pm~2.4 }$& $2.68~ {\pm~0.29 }$& $11.9 ~{\pm~0.3 }$& 46/41  \\
 B-04-02 & 11/15& 55150.31& $0.21~{\pm ~0.03}$& $6.05~{\pm ~0.02}$& $1.25~{\pm~ 0.14}$& $1.39~{\pm~ 0.08}$& 51/43 &$0.001~{\pm~2.1E-04  }$& $3.26~{\pm~0.58 }$& $225.2~{\pm~2.3 }$& $2.53~ {\pm~0.24 }$& $11.9 ~{\pm~0.3 }$& 40/41  \\
 B-06-00 & 01/19& 55215.91& $1.04~{\pm ~0.07}$& $9.86~{\pm ~0.07}$& $0.92~{\pm~ 0.03}$& $1.85~{\pm~ 0.06}$& 69/41 &$1.18 ~{\pm~0.12     }$& $5.40~{\pm~0.59 }$& $34.8~ {\pm~0.8 }$&  $3.94~{\pm~0.34 }$& $8.7  ~{\pm~0.2 }$& 38/38  \\
 B-06-01 & 01/20& 55216.95& $0.46~{\pm ~0.06}$& $8.09~{\pm ~1.03}$& $0.37~{\pm~ 0.02}$& $2.03~{\pm~ 0.07}$& 48/41 &$1.52 ~{\pm~0.24     }$& $6.12~{\pm~0.63 }$& $34.7~ {\pm~0.8 }$&  $3.99~{\pm~0.28 }$& $9.4  ~{\pm~0.3 }$& 42/38  \\
 B-06-02 & 01/21& 55217.87& $0.84~{\pm ~0.07}$& $7.03~{\pm ~0.24}$& $0.46~{\pm~ 0.09}$& $2.11~{\pm~ 0.08}$& 37/41 &$1.65 ~{\pm~0.22     }$& $5.75~{\pm~0.54 }$& $34.7~ {\pm~0.9 }$&  $3.99~{\pm~0.25 }$& $9.4  ~{\pm~0.2 }$& 42/38  \\
 C-01-08 & 01/22& 55218.14& $1.42~{\pm ~0.05}$& $7.81~{\pm ~0.57}$& $0.57~{\pm~ 0.06}$& $2.18~{\pm~ 0.01}$& 50/41 &$1.37 ~{\pm~0.21     }$& $5.45~{\pm~0.53 }$& $35.2~ {\pm~0.8 }$&  $3.90~{\pm~0.21 }$& $9.3  ~{\pm~0.2 }$& 47/38  \\
 C-01-00 & 01/22& 55218.80& $5.21~{\pm ~0.06}$& $4.38~{\pm ~0.20}$& $0.62~{\pm~ 0.05}$& $2.22~{\pm~ 0.01}$& 46/41 &$6.12 ~{\pm~0.80     }$& $1.87~{\pm~0.35 }$& $34.6~ {\pm~0.7 }$&  $1.07~{\pm~0.11 }$& $10.1  ~{\pm~0.2 }$& 66/37  \\
 C-01-02 & 01/24& 55220.68& $5.34~{\pm ~0.04}$& $3.50~{\pm ~0.30}$& $0.61~{\pm~ 0.04}$& $2.22~{\pm~ 0.01}$& 32/41 &$6.45 ~{\pm~0.70     }$& $2.21~{\pm~0.26 }$& $34.6~ {\pm~0.9 }$&  $1.16~{\pm~0.21 }$& $8.1  ~{\pm~0.3 }$& 55/37  \\
 C-01-10 & 01/25& 55221.35& $5.10~{\pm ~0.03}$& $2.07~{\pm ~0.03}$& $0.63~{\pm~ 0.04}$& $2.20~{\pm~ 0.03}$& 53/43 &$4.82 ~{\pm~0.40     }$& $1.60~{\pm~0.08 }$& $34.5~ {\pm~0.8 }$&  $1.27~{\pm~0.19 }$& $9.5  ~{\pm~0.3 }$& 56/38  \\
 C-01-12 & 01/26& 55222.33& $5.25~{\pm ~0.03}$& $0.97~{\pm ~0.03}$& $0.61~{\pm~ 0.05}$& $2.08~{\pm~ 0.07}$& 38/43 &$4.77 ~{\pm~0.45     }$& $1.26~{\pm~0.06 }$& $34.5~ {\pm~0.7 }$&  $1.21~{\pm~0.25 }$& $9.5  ~{\pm~0.2 }$& 47/38  \\
 C-01-14 & 01/28& 55224.36& $5.02~{\pm ~0.03}$& $1.27~{\pm ~0.03}$& $0.62~{\pm~ 0.04}$& $2.11~{\pm~ 0.05}$& 44/43 &$4.68 ~{\pm~0.43     }$& $1.03~{\pm~0.07 }$& $34.5~ {\pm~0.6 }$&  $1.15~{\pm~0.25 }$& $9.5  ~{\pm~0.2 }$& 43/38  \\
 C-02-02 & 01/30& 55226.25& $4.67~{\pm ~0.02}$& $1.70~{\pm ~0.02}$& $0.63~{\pm~ 0.04}$& $2.17~{\pm~ 0.03}$& 55/43 &$4.54 ~{\pm~0.43     }$& $0.91~{\pm~0.05 }$& $34.5~ {\pm~0.7 }$&  $1.10~{\pm~0.17 }$& $9.5  ~{\pm~0.2 }$& 54/38  \\
 C-03-00 & 02/05& 55232.98& $4.11~{\pm ~0.02}$& $0.89~{\pm ~0.01}$& $0.59~{\pm~ 0.04}$& $2.08~{\pm~ 0.04}$& 63/43 &$4.41 ~{\pm~0.43     }$& $0.72~{\pm~0.06 }$& $34.5~ {\pm~0.8 }$&  $1.10~{\pm~0.12 }$& $9.5  ~{\pm~0.3 }$& 58/38  \\
 C-06-00 & 02/26& 55253.51& $2.10~{\pm ~0.01}$& $0.27~{\pm ~0.01}$& $0.54~{\pm~ 0.04}$& $1.84~{\pm~ 0.09}$& 33/43 &$4.13 ~{\pm~0.45     }$& $0.58~{\pm~0.05 }$& $34.5~ {\pm~0.7 }$&  $1.12~{\pm~0.19 }$& $9.5  ~{\pm~0.3 }$& 40/38  \\
 C-09-04 & 03/23& 55278.58& $0.69~{\pm ~0.00}$& $0.16~{\pm ~0.00}$& $0.47~{\pm~ 0.06}$& $2.07~{\pm~ 0.11}$& 42/43 &$3.11 ~{\pm~0.21     }$& $0.52~{\pm~0.04 }$& $34.6~ {\pm~0.8 }$&  $1.10~{\pm~0.16 }$& $9.5  ~{\pm~0.2 }$& 36/37  \\
 C-09-05 & 03/24& 55279.63& $0.77~{\pm ~0.01}$& $0.16~{\pm ~0.00}$& $0.46~{\pm~ 0.06}$& $2.10~{\pm~ 0.12}$& 47/43 &$2.85 ~{\pm~0.22     }$& $0.54~{\pm~0.03 }$& $34.5~ {\pm~0.8 }$&  $1.07~{\pm~0.11 }$& $8.1  ~{\pm~0.2 }$& 41/37  \\
 C-10-01 & 03/27& 55282.57& $0.65~{\pm ~0.00}$& $0.20~{\pm ~0.00}$& $0.51~{\pm~ 0.07}$& $2.06~{\pm~ 0.06}$& 39/43 &$2.45 ~{\pm~0.28     }$& $0.60~{\pm~0.05 }$& $34.5~ {\pm~0.8 }$&  $1.06~{\pm~0.15 }$& $8.1  ~{\pm~0.2 }$& 25/37  \\
 C-10-04 & 03/30& 55285.44& $0.57~{\pm ~0.01}$& $0.69~{\pm ~0.01}$& $0.65~{\pm~ 0.02}$& $2.04~{\pm~ 0.03}$& 66/43 &$2.21 ~{\pm~0.24     }$& $0.66~{\pm~0.05 }$& $34.6~ {\pm~0.8 }$&  $1.05~{\pm~0.11 }$& $10.2 ~{\pm~0.3 }$& 32/37  \\
 C-11-01 & 04/03& 55289.36& $0.20~{\pm ~0.01}$& $1.64~{\pm ~0.01}$& $0.89~{\pm~ 0.09}$& $1.75~{\pm~ 0.02}$& 37/43 &$0.002~{\pm~2.0E-04  }$& $3.90~{\pm~0.30 }$& $126.1~{\pm~1.5 }$& $2.57~ {\pm~0.16 }$& $10.6 ~{\pm~0.3 }$& 69/41  \\
 C-11-05 & 04/08& 55294.26& $0.09~{\pm ~0.06}$& $1.53~{\pm ~0.01}$& $1.25~{\pm~ 0.12}$& $1.56~{\pm~ 0.04}$& 31/43 &$0.002~{\pm~2.3E-04  }$& $3.41~{\pm~0.37 }$& $127.6~{\pm~1.4 }$& $2.54~ {\pm~0.32 }$& $10.8 ~{\pm~0.4 }$& 30/41  \\
 C-12-03 & 04/13& 55299.95& $0.09~{\pm ~0.02}$& $1.19~{\pm ~0.02}$& $1.34~{\pm~ 0.13}$& $1.57~{\pm~ 0.04}$& 42/43 &$0.002~{\pm~2.3E-04  }$& $3.23~{\pm~0.40 }$& $128.6~{\pm~1.6 }$& $2.49~ {\pm~0.23 }$& $11.2 ~{\pm~0.2 }$& 42/41  \\
 C-12-04 & 04/15& 55301.80& $0.11~{\pm ~0.02}$& $1.11~{\pm ~0.02}$& $1.34~{\pm~ 0.15}$& $1.51~{\pm~ 0.05}$& 60/43 &$0.002~{\pm~2.4E-04  }$& $3.12~{\pm~0.41 }$& $128.0~{\pm~1.7 }$& $2.55~ {\pm~0.22 }$& $11.2 ~{\pm~0.3 }$& 34/41  \\
 D-01-03 & 04/19& 55305.58& $0.07~{\pm ~0.05}$& $0.94~{\pm ~0.49}$& $2.24~{\pm~ 0.18}$& $1.71~{\pm~ 0.03}$& 42/43 &$0.002~{\pm~2.4E-04  }$& $3.00~{\pm~0.40 }$& $129.2~{\pm~1.7 }$& $2.51~ {\pm~0.23 }$& $11.5 ~{\pm~0.4 }$& 42/41  \\
 D-02-01 & 04/24& 55310.70& $0.04~{\pm ~0.01}$& $0.84~{\pm ~0.01}$& $1.29~{\pm~ 0.17}$& $1.59~{\pm~ 0.05}$& 34/43 &$0.002~{\pm~2.2E-04  }$& $2.99~{\pm~0.32 }$& $128.9~{\pm~1.7 }$& $2.53~ {\pm~0.25 }$& $11.2 ~{\pm~0.3 }$& 35/41  \\
 D-02-03 & 04/26& 55312.60& $0.07~{\pm ~0.03}$& $0.76~{\pm ~0.03}$& $0.63~{\pm~ 0.18}$& $1.62~{\pm~ 0.03}$& 41/43 &$0.002~{\pm~2.1E-04  }$& $2.92~{\pm~0.37 }$& $130.4~{\pm~1.2 }$& $2.49~ {\pm~0.23 }$& $11.5 ~{\pm~0.3 }$& 44/41  \\
 D-03-00 & 04/30& 55316.05& $0.08~{\pm ~0.02}$& $0.66~{\pm ~0.02}$& $1.07~{\pm~ 0.17}$& $1.47~{\pm~ 0.07}$& 41/43 &$0.002~{\pm~2.2E-04  }$& $2.90~{\pm~0.33 }$& $129.5~{\pm~1.9 }$& $2.60~ {\pm~0.26 }$& $9.5  ~{\pm~0.3 }$& 46/41  \\
 D-03-02 & 05/02& 55318.55& $0.04~{\pm ~0.01}$& $0.63~{\pm ~0.01}$& $0.34~{\pm~ 0.17}$& $1.68~{\pm~ 0.05}$& 41/43 &$0.002~{\pm~2.3E-04  }$& $2.85~{\pm~0.34 }$& $129.3~{\pm~1.4 }$& $2.59~ {\pm~0.19 }$& $9.4  ~{\pm~0.3 }$& 39/41  \\
 D-04-01 & 05/08& 55325.00& $0.04~{\pm ~0.04}$& $0.40~{\pm ~0.03}$& $1.89~{\pm~ 0.31}$& $1.54~{\pm~ 0.13}$& 30/43 &$0.002~{\pm~1.7E-04  }$& $2.70~{\pm~0.39 }$& $133.6~{\pm~1.6 }$& $2.57~ {\pm~0.22 }$& $11.2 ~{\pm~0.3 }$& 33/41  \\
 D-08-02 & 06/09& 55356.17& $0.02~{\pm ~0.04}$& $0.36~{\pm ~0.02}$& $1.13~{\pm~ 0.62}$& $1.56~{\pm~ 0.11}$& 36/43 &$0.002~{\pm~3.1E-04  }$& $2.63~{\pm~0.41 }$& $129.9~{\pm~1.8 }$& $2.65~ {\pm~0.30 }$& $9.4  ~{\pm~0.3 }$& 35/41  \\
 D-10-01 & 06/21& 55368.94& $0.02~{\pm ~0.02}$& $0.25~{\pm ~0.01}$& $1.26~{\pm~ 0.31}$& $1.47~{\pm~ 0.13}$& 33/43 &$0.002~{\pm~2.8E-04  }$& $2.54~{\pm~0.33 }$& $129.6~{\pm~1.5 }$& $2.56~ {\pm~0.29 }$& $9.4  ~{\pm~0.4 }$& 33/41  \\

 \hline
 \end{tabular}
 \noindent{
 \leftline{$^{[1]}$ A, B, C, D mark initial part of the observation IDs: 94044-07, 94331-01, 95360-01, 95702-01 respectively.} 
 \leftline{$^{[2]}$ UT dates are in mm/dd format. First 4 observations are from 2009 and rest from 2010.}
 \leftline{$^{[3]}$ $DBB+PL$ model fitted individual model fluxes (in $10^{-9}~erg~cm^{-2}~s^{-1}$) and parameters disk : temperature $T_{in}$ (in keV),}
 \leftline{and PL index ($\Gamma$) are mentioned in Cols. 4-7.}
 \leftline{$^{[4]}$ TCAF model fitted parameters: disk rate (${\dot m}_d$ in Eddington rate ${\dot M}_{Edd}$), halo rate (${\dot m}_h$ in ${\dot M}_{Edd}$), shock location ($X_s$),} 
 \leftline{compression ratio ($R$), mass of the black hole ($M_{BH}$ in solar mass $M_{\odot}$) are mentioned in Cols. 9-13.}
 \leftline{$^{[5]}$ $DBB+PL$ model or TCAF model fitted ${\chi}^2_{red}$ values are mentioned in column 8 and 14 respectively as ${\chi}^2/dof$,} 
 \leftline{where `dof' represents degrees of freedom.}
\leftline {Note: we present average values of 90\% confidence $\pm$ parameter error values, which are obtained using `err' task in XSPEC.}          }
\end{table*}

\begin{table*}
 \addtolength{\tabcolsep}{-0.5pt}
 \centering
 \caption{Fractional Disk Flux and Integrated Power Continuum}
 \label{tab}
 \begin{tabular}{|c|c|c|c|c|c|c|c|}
 \hline

Obs ID$^{[1]}$ &  UT$^{[2]}$ & MJD$^{[2]}$ &    disk-frac (\%)$^{[3]}$ &  $r^{[4]}$ &       $\nu^{[5]}$   &    rms (\%)$^{[5]}$   &  QPO type$^{[5]}$\\
   1   &  2  &  3  &         4             &          5            &       6        &     7      &      8      \\
\hline

 A-01-00  &  10/26  &  55130.93   &  4.39   &  0.12      &       &          &   \\
 B-02-06  &  11/02  &  55137.23   &  5.55   &  0.12      &       &          &   \\
 B-02-10  &  11/04  &  55139.58   &  4.20   &  0.12      &       &          &   \\
 B-04-02  &  11/15  &  55150.31   &  4.53   &  0.11      &       &          &   \\
 B-06-00  &  01/19  &  55215.91   &  12.6   &  5.0E-02   &  2.20 & 10.95    & C \\
 B-06-01  &  01/20  &  55216.95   &  9.65   &  3.3E-02   &  4.02 & 10.53    & C \\
 B-06-02  &  01/21  &  55217.87   &  13.6   &  2.3E-02   &  5.29 & 8.87     & C \\
 C-01-08  &  01/22  &  55218.14   &  25.2   &  8.9E-03   &  6.15 & 4.73     & C \\
 C-01-00  &  01/22  &  55218.80   &  42.3   &  6.3E-03   &  3.51 & 7.15     & B \\
 C-01-02  &  01/24  &  55220.68   &  46.4   &  2.5E-03   &  2.19 & 4.50     & B \\
 C-01-10  &  01/25  &  55221.35   &  77.8   &  4.6E-03   &       &          &   \\
 C-01-12  &  01/26  &  55222.33   &  88.8   &  6.6E-04   &       &          &   \\
 C-01-14  &  01/28  &  55224.36   &  85.1   &  2.4E-03   &       &          &   \\
 C-02-02  &  01/30  &  55226.25   &  79.8   &  2.9E-03   &       &          &   \\
 C-03-00  &  02/05  &  55232.98   &  87.3   &  4.0E-04   &       &          &   \\
 C-06-00  &  02/26  &  55253.51   &  92.3   &  1.4E-03   &       &          &   \\
 C-09-04  &  03/23  &  55278.58   &  87.6   &  9.2E-04   &       &          &   \\
 C-09-05  &  03/24  &  55279.63   &  84.7   &  2.4E-03   &       &          &   \\
 C-10-01  &  03/27  &  55282.57   &  69.7   &  7.8E-03   &       &          &   \\
 C-10-04  &  03/30  &  55285.44   &  28.9   &  2.9E-02   &       &          &   \\
 C-11-01  &  04/03  &  55289.36   &  14.8   &  4.6E-02   &       &          &   \\
 C-11-05  &  04/08  &  55294.26   &  7.34   &  5.7E-02   &       &          &   \\
 C-12-03  &  04/13  &  55299.95   &  9.22   &  5.4E-02   &       &          &   \\
 C-12-04  &  04/15  &  55301.80   &  12.0   &  5.9E-02   &       &          &   \\
 D-01-03  &  04/19  &  55305.58   &  9.34   &  4.7E-02   &       &          &   \\
 D-02-01  &  04/24  &  55310.70   &  7.11   &  4.8E-02   &       &          &   \\
 D-02-03  &  04/26  &  55312.60   &  10.4   &  4.6E-02   &       &          &   \\
 D-03-00  &  04/30  &  55316.05   &  13.6   &  4.1E-02   &       &          &   \\
 D-03-02  &  05/02  &  55318.55   &  8.68   &  4.7E-02   &       &          &   \\
 D-04-01  &  05/08  &  55325.00   &  10.5   &  4.3E-02   &       &          &   \\
 D-08-02  &  06/09  &  55356.17   &  6.61   &  3.9E-02   &       &          &   \\
 D-10-01  &  06/21  &  55368.94   &  13.0   &  3.3E-02   &       &          &   \\

\hline
 \end{tabular}

\noindent{
\leftline{$^{[1]}$ A, B, C, D mark initial part of the observation IDs: 94044-07, 94331-01, 95360-01, 95702-01 respectively.}
 \leftline{$^{[2]}$ UT dates (in mm/dd format) and modified Julian day of observations. First 4 observations are from 2009 and rest from 2010.}
 \leftline{$^{[3]}$ Percentage of the soft (DBB) flux with the total (DBB+PL) flux.}
 \leftline{$^{[4]}$ $r$ is the integrated power continuum in $0.1 - 10$ Hz range of PDS.} 
 \leftline{$^{[5]}$ QPO properties are mentioned,. Here $\nu$ is the observed QPO frequency.}
}
\end{table*}

\end{document}